# Strong-coupling limit in cold-molecule formation via photoassociation or Feshbach resonance through Nikitin exponential resonance crossing

## Artur Ishkhanyan[1] and Hiroki Nakamura[2]

[1]*Engineering Center of NAS of Armenia, Ashtarak-2, 0203 Armenia*
[2]*Institute for Molecular Science, Okazaki 444-8585, Japan*

The strong-coupling limit of molecule formation in an atomic Bose-Einstein condensate via two-mode one-color photoassociation or sweep across a Feshbach resonance is examined using a basic nonlinear time-dependent two-state model. For the general class of term-crossing models with constant coupling, a common strategy for attacking the problem is developed based on the reduction of the initial system of semiclassical equations for atom-molecule amplitudes to a third order nonlinear differential equation for the molecular state probability. This equation provides deriving *exact* solution for a class of *periodic* level-crossing models. These models reveal much in common with the Rabi problem. Discussing the strong-coupling limit for the general case of variable detuning, the equation is further truncated to a limit first-order nonlinear equation. Using this equation, the strong nonlinearity regime for the first Nikitin exponential-crossing model is analyzed and accurate asymptotic expressions for the nonlinear transition probability to the molecular state are derived. It is shown that, because of a *finite final detuning* involved, this model displays essential deviations from the Landau-Zener behavior. In particular, it is shown that in the limit of strong coupling the final conversion probability tends to 1/6. Thus, in this case the strong interaction limit is not optimal for molecule formation. We have found that if optimal field intensity is applied the molecular probability is increased up to 1/4 (i.e., the half of the initial atomic population).

**1. Introduction**

Two major techniques currently widely used for molecule production from cold atoms are the photoassociation reaction [1] and the Feshbach resonance [2]. In both cases, a sweep through a resonance plays a crucial role to achieve significant molecular population. Being well known for a long time (Landau and Zener introduced a prototypical concept as early as in 1932 [3]), this observation caused constant interest to the term-crossing models describing corresponding resonance processes and currently several such models are developed to embrace different aspects of associated nonadiabatic transitions [4-10]. However, these models deal with linear processes while the recent extensive developments in the physics of Bose-Einstein condensates and degenerate Fermi gases demand consideration of term crossings in nonlinear systems. The nonlinearity (it here stems from many-body effects) is potent to drastically change the interaction picture [11]. (This is why a renewal of interest in the term-crossing problems and, as a result, a notable research activity, both theoretical and experimental, is observed towards development of nonlinear term-crossing models during the past years [11-18].) Of course, the changes should be more expressed in the strong-coupling regime of high interaction strengths, when the nonlinearity is well pronounced. In the present paper we examine this regime using different term-crossing models. A motivation for this research is that there is a gap between the actual form of the sweeping through the resonance applied in the recent experiments (see, e.g., [15-18]) and the Landau-Zener *linear-crossing* model used so far in the theoretical approaches (e.g., [12-14]) to interpret the obtained experimental results. The analysis of the current situation shows that more realistic models with different properties should be applied. For this reason we develop the first Nikitin exponential-crossing model that is close to the Landau-Zener one at the vicinity of the crossing but involves a *finite final detuning*. This detuning presents an additional parameter, actually present in the experiment, and hence this model is expected to be more appropriate to interpret the experimental observations. We show that the model indeed displays essential deviations from the Landau-Zener behavior. In particular, it turns out that in contrast to the Landau-Zener case the strong coupling is not the best regime for molecule production.

In our development we use the simplest, semiclassical coupled two-mode approach when the two processes, photoassociation and Feshbach resonance, are described by the same set of first-order nonlinear time-dependent equations treating the atomic and molecular populations as classical fields. In order to deal with familiar quantum optics notations and be in the position to use corresponding analogies as well as accumulated knowledge concerning quantum nonadiabatic transitions, we use throughout the photoassociation terminology.

The system of coupled *nonlinear* equations governing the time evolution of an effective two-state quantum system under consideration is written as [1,2,19,20]

$$i\frac{da_1}{dt} = U(t)e^{-i\delta(t)}\bar{a}_1 a_2, \quad i\frac{da_2}{dt} = \frac{U(t)}{2}e^{i\delta(t)}a_1 a_1. \quad (1)$$

Here $a_1$ and $a_2$ are the atomic and molecular states' amplitudes, respectively, $\bar{a}_1$ is the complex conjugate to $a_1$, $U(t)$ is the Rabi frequency associated with the photoassociating laser field amplitude, and $\delta(t)$ is the corresponding phase modulation function whose derivative $\delta_t = d\delta/dt$ is the laser field frequency detuning from the transition frequency. All the quantities involved are supposed to be dimensionless. The initial conditions considered here are $|a_1(-\infty)|^2 = 1$ and $|a_2(-\infty)|^2 = 0$. System (1) preserves the number of particles that we normalize to unity: $|a_1|^2 + 2|a_2|^2 = \text{const} = 1$. It is not difficult to show that with this normalization, the probability for the molecular state, $p(t) = |a_2(t)|^2$, obeys the following nonlinear ordinary differential equation of the third order:

$$p_{ttt} - \left(\frac{\delta_{tt}}{\delta_t} + 2\frac{U_t}{U}\right)p_{tt} + \left[\delta_t^2 + 4U^2(1-3p) - \left(\frac{U_t}{U}\right)_t + \frac{U_t}{U}\left(\frac{\delta_{tt}}{\delta_t} + \frac{U_t}{U}\right)\right]p_t$$
$$+ \frac{U^2}{2}\left(\frac{\delta_{tt}}{\delta_t} - \frac{U_t}{U}\right)(1 - 8p + 12p^2) = 0, \qquad (2)$$

where (and hereafter) the alphabetical subscripts denote differentiation. In the case of constant field amplitude, $U = U_0 = \text{const}$, this equation is significantly simplified:

$$p_{ttt} - \frac{\delta_{tt}}{\delta_t}p_{tt} + \left[\delta_t^2 + 4U_0^2(1-3p)\right]p_t + \frac{U_0^2}{2}\frac{\delta_{tt}}{\delta_t}(1 - 8p + 12p^2) = 0. \qquad (3)$$

We have previously shown [21] that this equation is equivalent to a nonlinear Volterra integral equation of the second kind [22]. The latter allows one to construct uniformly convergent series solution for the case of small $U_0^2$ using Picard's successive approximations [22]. We have used this approach in treating the Landau–Zener problem [3]. Other analogous models can be straightforwardly considered using this Volterra integral equation. Thus, the method can be adopted as a general strategy for attacking the nonlinear two-state problems in the limit of weak coupling, $U_0^2 \ll 1$. However, the opposite limit of ***strong*** field intensities, $U_0^2 \gg 1$, presents a much more difficult problem. And the investigation of this limit is, as was already said above, the primary task of the present research. We analyze and compare different level-crossing models and derive exact or approximate formulas for the transition probability. The evolution of the molecular state probability as a function of time displays significant anomalies as compared with the linear case. We show that for some field configurations (e.g., for the first Nikitin exponential model discussed below) the strong coupling limit is not optimal for conversion of atoms into the molecular state. Such a behavior, indeed, substantially differs from the linear two-level system's response to strong laser field excitation.

Our treatment of the strong-coupling limit is essentially based on Eq. (3) where in this case we have a large parameter, the field intensity $U_0^2$. Brief examination of the equation then suggests that the most important terms, probably, are the last two, i.e., one may try to construct an initial approximation by ignoring the first two terms with second- and third-order derivatives. Though neglecting the higher-order derivatives is a singular procedure in general, for our particular equation, as we show below, this works: at high field intensities the behavior of the system in the most cases is effectively governed by the first-order nonlinear equation formed by the two last terms of Eq. (3). The Rabi problem which assumes constant frequency detuning $(\delta_t = \mathrm{const})$ is a noteworthy exception since then, as it is immediately seen, the last term of the equation (as well as the second one) identically vanishes. Therefore, in this case we have to consider a third-order nonlinear equation. As a result, the structure of the solution is significantly modified. Fortunately, the Rabi problem is treated exactly, without approximations (see, e.g., [23]). This is a useful point since some models, such as the Nikitin exponential [5] and the second Demkov-Kunike [6] model, include large time regions where the detuning is practically constant. It is then expected that these models reveal features that are generic for the Rabi problem. We will convince the reader below that this is, indeed, the case. We will also see that some features common for the Rabi problem are well pronounced in the periodic level-crossing models that we present.

For the above reasons, we first briefly review the Rabi problem and discuss some related periodic term-crossing models. Further, we examine the general case of variable detuning and develop a common approach applicable to all models. Using the approach, we derive a simple approximate formula for the final probability of the transition into the molecular state for the case of the first Nikitin exponential-crossing model [5]. Further, we examine this case in detail and show that for optimal conversion into the molecular state the applied field intensity should be adjusted depending on the corresponding frequency detuning.

## 2. The Rabi problem and periodic term-crossing models

The Rabi problem is characterized by a constant detuning, $\delta_t = \delta_0$ (recall that the field amplitude is a constant as well, $U = U_0$). Eq. (3) takes the form

$$p_{ttt} + [\delta_0^2 + 4U_0^2(1-3p)]p_t = 0. \tag{4}$$

This equation is readily integrated once to give

$$p_{tt} + [(\delta_0^2 + 4U_0^2)p - 6U_0^2 p^2] = C_1/2. \tag{5}$$

Then, by putting $p_t = \Phi(p)$, we obtain

$$\Phi^2 + [(\delta_0^2 + 4U_0^2)p^2 - 4U_0^2 p^3] = C_1 p + C_0, \tag{6}$$

so that we arrive at

$$t - t_0 = \int \frac{dp}{\sqrt{C_0 + C_1 p - (\delta_0^2 + 4U_0^2)p^2 + 4U_0^2 p^3}}, \tag{7}$$

where $C_0 = 0$ and $C_1 = U_0^2$, according to the initial conditions $p(t_0) = p_t(t_0) = 0$, $p_{tt}(t_0) = U_0^2/2$. The integral involved here is reduced to an elliptic integral of the first kind so that the solution is finally written in terms of the Jacobi elliptic function [24]:

$$p = p_1 \, \text{sn}^2[\sqrt{p_2} U_0 (t - t_0); m], \tag{8}$$

$$p_{1,2} = \frac{1}{2}\left(\frac{\delta_0^2}{4U_0^2} + 1\right) \mp \sqrt{\frac{1}{4}\left(\frac{\delta_0^2}{4U_0^2} + 1\right)^2 - \frac{1}{4}}, \quad m = \frac{p_1}{p_2}. \tag{9}$$

In general, this is a periodic solution with the period given as

$$T(m) = \frac{\pi}{\sqrt{p_2} U_0} \cdot {}_2F_1(1/2, 1/2; 1; m), \tag{10}$$

where ${}_2F_1$ is the Gauss hypergeometric function [24].

At weak interaction regime $m \approx 0$ and we have slightly perturbed linear Rabi sinusoidal oscillations. However, the field intensity being increased, more and more harmonics emerge and the shape of the function becomes more rectangular with increasing length (Fig. 1). Eventually, in the strong coupling limit, $U_0^2 \to \infty$ ($m \to 1$), we have

$$p = \frac{1}{2}\tanh^2\left[\frac{U_0(t - t_0)}{\sqrt{2}}\right], \tag{11}$$

i.e., total transition to the molecular state at $t \to \infty$ when starting from a pure atomic condensate (Fig. 1). Evidently, this limit, $U_0^2 \to \infty$, is equivalent to the case of exact resonance at finite field intensities: $U_0^2 < \infty$.

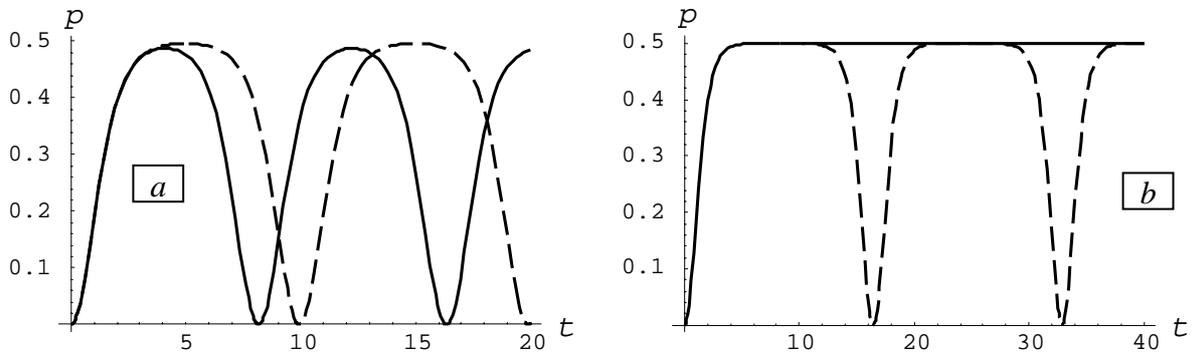

Fig. 1. Rabi solution:
(a) intermediate regime, $U_0^2/\delta_0^2 \sim 1$, (b) strong coupling limit, $U_0^2/\delta_0^2 \gg 1$.

A related ***exactly solvable*** class of periodic-crossing models is derived by putting $\delta_t = f(p)$ [or, more generally, $U = U(t)$ and $\delta_t = U(t)f(p)$]. This can be easily seen by rewriting Eq. (2) in an equivalent form:

$$\left(\frac{p_{tt}}{\delta_t}\right)_t + \delta_t p_t - \frac{U_0^2}{2}\left(\frac{1-8p+12p^2}{\delta_t}\right)_t = 0. \tag{12}$$

It is seen that for $\delta_t = \delta_t(p)$ the equation is integrated to give

$$p_{tt} - \frac{U_0^2}{2}(1-8p+12p^2) + \delta_t(p)\int_{t_0}^{t}\delta_t(p)dp = C_1\delta_t(p), \quad C_1 = \text{const}, \tag{13}$$

where should be $C_1 = 0$ to fulfill the initial conditions. This leads to the solution of the form [compare with (7)]

$$t - t_0 = \int\frac{dp}{\sqrt{-2Q(p) + U_0^2(p - 4p^2 + 4p^3)}}, \quad Q(p) = \int_{t_0}^{t}\delta_t(p)\left(\int_{t_0}^{t}\delta_t(p)\,dp\right)dp. \tag{14}$$

As is known, if $Q(p)$ is a cubic or quartic polynomial in $p$, the solution $p(t)$ is written in terms of elliptic functions [24] and thus represents, in general, a periodic function of time. Hence, the corresponding detuning $\delta_t = \delta_t(p)$ defines a ***periodic term-crossing*** model. A number of such periodic-crossing models can be derived by choosing different functions $Q(p)$. As it is immediately seen from Eqs. (14) and (7), these models have much in common with the above Rabi solution. {In a certain sense, the Rabi model can also be included in the list under consideration as a particular limiting case [it corresponds to $Q(p) = \delta_0^2 p^2/2$]}.

For instance, if

$$U = U_0 = \text{const}, \quad \delta_t = \delta_1\sqrt{p}, \tag{15}$$

then $Q(p) = 2\delta_1^2 p^3/9$ and the solution is given by

$$p = b\,\text{sn}^2\left[\frac{1}{2\sqrt{b}}U_0(t-t_0); m\right], \tag{16}$$

where
$$m = -\frac{b}{a} = \frac{3-\delta_1/U_0}{3+\delta_1/U_0}, \quad a,b = \frac{3}{2(\delta_1/U_0 \mp 3)}. \tag{17}$$

Interestingly, at $\delta_1 = 3U_0$ ($m = 0$) $\Leftrightarrow \delta_t = 3U_0\sin[U_0(t-t_0)]$ the solution is expressed in terms of elementary functions: $p = \sin^2[U_0(t-t_0)/4]$. As is seen, this solution, indeed, has the same structure as the Rabi solution (8). And in the strong coupling limit we have the same limit function (11). However, it should be said here that the presented periodic-crossing models are rather degenerate since the frequency and amplitude of the photoassociating laser field are not independent. Nevertheless, they can be useful in constructing perturbative solutions to more realistic nondegenerate problems.

**3. General case of variable detuning: limit equation**

Consider now the strong interaction limit in the general case when $\delta_t$ is not constant so that $\delta_{tt} \neq 0$. Taking into account the orders of the involved terms, we first keep in Eq. (3) only the last two terms:

$$\left[\delta_t^2 + 4U_0^2(1-3p)\right]p_t + \frac{U_0^2}{2}\frac{\delta_{tt}}{\delta_t}\left(1-8p+12p^2\right) = 0. \tag{18}$$

Though simple at first glance, this equation has a rich structure. For all the models, it possesses two trivial solutions: $p = 1/6$ and $p = 1/2$. Notably, these are the stationary solutions of the exact equation (3). These solutions play an important role in establishment of the asymptotes. Besides, Eq. (18) has a remarkable nontrivial solution. The general solution depending on a constant $C$ has a rather complicated structure. This solution is considerably simplified at two specific choices of the constant. In the case of the Landau-Zener model [3], $\delta = \delta_0 t^2$, the particular solutions generated by these choices of $C$ are written as:

$$p(t) = \frac{1}{6} + \frac{2t}{9\lambda}\left(t \pm \sqrt{t^2 + \frac{3\lambda}{2}}\right) \text{ and } p(t) = \frac{1}{2} + \frac{2t}{9\lambda}\left(t \pm \sqrt{t^2 - \frac{3\lambda}{2}}\right), \tag{19}$$

where $\lambda = U_0^2/\delta_0$. Note that the last two functions are not defined on the whole real axes. Since none of remaining two solutions is bounded, it is understood that none of functions (19) can define the approximate solution to the source equation (3) alone. The normalization constraint imposes further restrictions to the applicability regions of these functions. Now, the analysis shows that in the strong coupling regime the limit solution at $\lambda \to \infty$ to the Landau-Zener problem subject to the initial conditions considered here is composed from pieces of different solutions, namely, the nontrivial solution

$$p_0(t) = \frac{1}{6} + \frac{2t}{9\lambda}\left(t + \sqrt{t^2 + \frac{3\lambda}{2}}\right) \quad \text{when} \quad t < \sqrt{\frac{\lambda}{2}} \tag{20}$$

and the trivial solution $\quad p_0(t) = \frac{1}{2} \quad$ when $\quad t > \sqrt{\frac{\lambda}{2}}$. $\tag{21}$

This *composite* solution is rather good approximation everywhere except a small region of the point $t = \sqrt{\lambda/2}$ where, additionally, discontinuity in derivatives is encountered. Furthermore, importantly, this limit solution allows one to linearize the initial problem (3) using the substitution $p = p_0 + u$. In this way, one arrives at a remarkable formula for the Landau-Zener transition probability [25], stating that in the strong coupling limit the transition probability is a linear function of the resonance crossing rate [25] (see also [13,14]), as opposed to the exponential dependence known for the counterpart linear problem [3].

Thus the limit equation (18) may provide essential advance. We have checked that this is the case for all familiar two-state models for nonadiabatic transitions such as the Rosen-Zener [4], first and second Demkov-Kunike [6], first [5] and second [7] Nikitin, Bambini-Berman [8] models, etc. (a rigorous general treatment is presented in a different paper [26]). Due to its rich structure, this equation gives a comprehensive general insight about the acting mechanisms and resultant physical processes at high field intensities. In particular, it leads to a general conclusion of both theoretical and practical importance that the molecular state probability is always very close to $1/6$ at the resonance crossing point and that no noncrossing model is able to provide final transition probability exceeding $1/6$ [26]. Below we demonstrate the potential of Eq. (18) using the first Nikitin exponential level-crossing model.

## 4. The first Nikitin exponential model

Let the photoassociating field amplitude be constant, hence, $U = U_0$, and consider the following particular detuning modulation function:

$$\delta_t = \Delta(1 - e^{-at}). \qquad (22)$$

This three-parametric *field configuration* (Fig. 2) is known as the first Nikitin exponential term-crossing *model* [5]. (One should distinguish this constant-amplitude model from the second Nikitin exponential model [7] where the field amplitude is also an exponential function: $U = U_0 e^{-at}$; the latter model is referred to as standard Nikitin model.) The set of *linear* governing equations [compare with Eq. (1)]

$$i\frac{da_{1L}}{dt} = U(t)e^{-i\delta(t)}a_{2L}, \quad i\frac{da_{2L}}{dt} = U(t)e^{i\delta(t)}a_{1L} \qquad (23)$$

with this field configuration forms the linear Nikitin two-state *problem* often faced in the theory of quantum nonadiabatic transitions (see, e.g., [7,27-29]). This problem models a rather flexible situation due to the fact that it includes the curve crossing (Landau-Zener) and noncrossing (Rosen-Zener) processes as limiting cases and therefore provides a description of other possibilities, for instance, broad avoided crossings in the theory of atomic collisions. Being frequently a subject of theoretical interest in several physical contexts, the linear

Nikitin exponential problem has undergone considerable development (e.g., towards solution of the time-independent quantum-mechanical two-state exponential problem [28]) after the determination of the nonadiabatic transition probabilities, under specific initial conditions, in the first works of Nikitin. Regarding the counterpart nonlinear two-state problem discussed here, so far there are no known analogous developments.

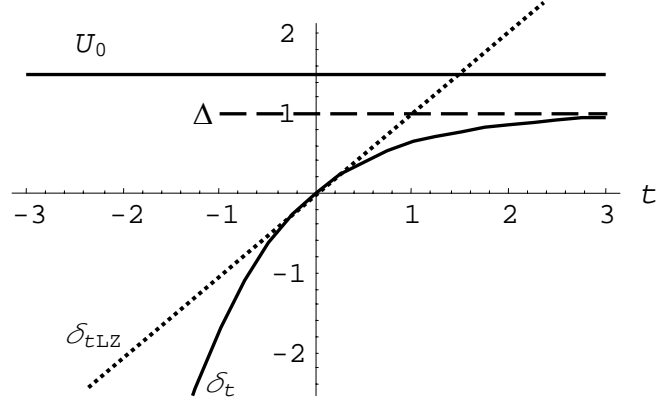

Fig. 2. The first Nikitin exponential term-crossing model ($U_0 = 1.5$, $\Delta = 1$, $a > 0$).

In the detuning function (22), the term crossing point is adjusted to coincide with the origin. As is seen (Fig. 2), at the vicinity of this point the detuning behaves similarly to the Landau–Zener linear function (shown in Fig. 2 by a dotted line), $\delta_{tLZ} = \Delta a t$. (As an effective Landau–Zener parameter, here the parameter $\lambda_{LZ} = 2U_0^2 / |a\Delta|$ stands). It is for this reason that the Nikitin model includes the features of the Landau-Zener model. On the other hand, note that at $t \gg 1/a$ ($a > 0$) the detuning is practically constant. Therefore, one may expect that this model will also incorporate the characteristics of the Rabi problem. The analysis below shows that this is indeed the case. It turns out that while in the weak interaction and extreme strong coupling limits the Rabi-type evolution, i.e., the Rabi-type periodic oscillations, is vaguely expressed, in the intermediate regime of moderate field intensities well pronounced, large amplitude Rabi oscillations are observed. This leads to essential differences as compared with the Landau-Zener transition already in the linear case. In particular, it turns out that at strong coupling only the half of the population undergoes transition to the second energy level, while in the Landau-Zener case all the population is changed to the second level. Furthermore, the differences become more expressed in the nonlinear case. For instance, we show that now only the third of the initial atomic population is capable to change into the molecular state.

The general solution of the linear problem (23) for second level's amplitude is written in terms of the Kummer confluent hypergeometric functions [24] as follows:

$$a_{2L} = C_1 e^{-\alpha_1 a t} {}_1F_1(\alpha_1; \gamma_1; i\Delta e^{-at}/a) + C_2 e^{-\alpha_2 a t} {}_1F_1(\alpha_2; \gamma_2; i\Delta e^{-at}/a), \quad (24)$$

where
$$\alpha_{1,2} = -i\frac{\Delta \mp R}{2a}, \quad \gamma_{1,2} = 1 \pm i\frac{R}{a} \quad (25)$$

and $R = \sqrt{\Delta^2 + 4U_0^2}$ is the effective Rabi frequency. The probability amplitude of the first level is then given according to Eqs. (22) and (23) as

$$a_{1L} = \frac{i}{U_0 e^{i\Delta(t+e^{-at}/a)}} \frac{da_{2L}}{dt}. \quad (26)$$

The system behavior defined by this solution depends on several factors. First of all, it depends on the sign of $a$, that is, effectively, the direction of the resonance crossing. Indeed, at $a > 0$ the detuning starts from infinity at $t \to -\infty$ and reaches the finite value $\Delta$ at $t \to +\infty$, while at $a < 0$ the detuning is initially finite and diverges at the end of the process. Obviously, these two situations are highly asymmetric. The process adequate to that described by the Landau-Zener model is the one corresponding to positive $a$. Furthermore, the transition probability is rather sensitive to the initial conditions. Note that the cases $a > 0$ and $a < 0$ assume essentially different types of initial conditions. For instance, Eq. (24) shows that at $a < 0$ the initial conditions should be, necessarily, oscillatory in time: $a_{2L}(t \to -\infty) \sim C_1 e^{-\alpha_1 a t} + C_2 e^{-\alpha_2 a t}$, while at $a > 0$ the system may start from the first energy level. Assuming thus $a > 0$ and applying initial conditions $a_{1L}(-\infty) = 1$, $a_{2L}(-\infty) = 0$ we obtain the constants $C_{1,2}$

$$C_1 = \frac{U_0}{R}\left(\frac{|\Delta|}{a}\right)^{-\alpha_2} \frac{\Gamma(\gamma_2)}{\Gamma(1-\alpha_1)} e^{S\pi\frac{\Delta+R}{4a}}, \quad C_2 = -\frac{U_0}{R}\left(\frac{|\Delta|}{a}\right)^{-\alpha_1} \frac{\Gamma(\gamma_1)}{\Gamma(1-\alpha_2)} e^{S\pi\frac{\Delta-R}{4a}}, \quad (27)$$

where $\Gamma$ is the Euler gamma-function [24] and $S = \text{sign}(\Delta)$. Note that the initial conditions applied here assume the normalization $|a_{1L}|^2 + |a_{2L}|^2 = 1$.

Finally, the transition probability depends on the specific finite state that is of interest for the particular physical problem under consideration. Originally, Nikitin calculated the probabilities of the transition to the *quasienergy* states (i.e., states corresponding to a certain Floquet characteristic exponent, see [30]). For the first such state defined as

$$\begin{Bmatrix} a_1(+\infty) \\ a_2(+\infty) \end{Bmatrix} \sim \begin{Bmatrix} \dfrac{i}{U_0 e^{i\Delta t}}(-\alpha_1 a)C_1 \\ C_1 \end{Bmatrix} e^{-\alpha_1 a t} \tag{28}$$

Nikitin's result reads [5]

$$p_{\to I} = \frac{1-e^{\pi(\Delta-R)/a}}{1-e^{-2\pi R/a}}. \tag{29}$$

This expression demonstrates the close relation between the Nikitin and Landau-Zener models. Indeed, the numerator of this expression agrees with the Landau-Zener formula (with $\lambda_{LZ} = 2U_0^2/|a\Delta|$) at $\Delta \to \infty$, and the denominator, which approaches unity at $\Delta \to \infty$, is the correction accounting for the involved finite detuning ("finite splitting between terms at infinity", if original collision terminology of [5] is used).

However, note that we are here interested in a different characteristic quantity, namely, the transition probability to the second energy level, i.e., $p_L = |a_{2L}|^2$ - this is the quantity corresponding to the molecule formation probability in the counterpart nonlinear case. The asymptotic solution for this probability is very sensitive to the type of asymptotics applied (see, e.g., [29]). However, for our purposes it is sufficient to apply the following approximation which is valid for any set of input field parameters $U_0, \Delta, a$ if sufficiently long interaction times are considered

$$p_L \sim p_{0L} + A\cos\left(-Rt + \frac{R}{a}\ln\frac{|\Delta|}{a} + \arg(\Phi)\right), \tag{30}$$

where

$$p_{0L} = \frac{1}{2} + \frac{|\Delta|}{2R}\frac{\cosh(\pi R/a) - \exp(\pi|\Delta|/a)}{\sinh(\pi R/a)}, \tag{31}$$

$$A = \frac{2U_0}{R}\frac{e^{\pi|\Delta|/(2a)}}{\sinh(\pi R/a)}\sqrt{\sinh\left(\frac{\pi(R-\Delta)}{2a}\right)\sinh\left(\frac{\pi(R+\Delta)}{2a}\right)}, \tag{32}$$

and

$$\Phi = \frac{\Gamma(\alpha_1)\Gamma(-\alpha_2)}{\Gamma^2(\gamma_1)}. \tag{33}$$

As it is immediately seen from Eq. (30), the final transition probability displays well expressed periodic Rabi oscillations. However, as is seen from Fig. 3, at high field intensities the amplitude of the oscillations is considerably decreased, so that the transition probability at strong-coupling limit is effectively defined by the nonoscillatory term $p_{0L}$ (average transition probability).

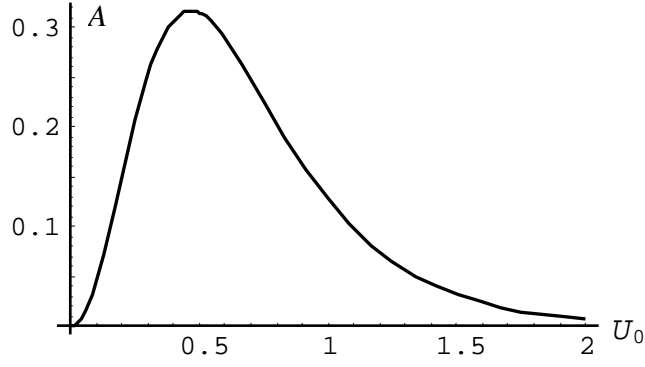

Fig. 3. The amplitude of Rabi oscillations vs Rabi frequency ($\Delta = 1$, $a = 1$), Eq (32).

As a result, for the strong coupling limit we finally obtain that the final transition probability at $t \to +\infty$ behaves as

$$p_L \sim \frac{1}{2} + \frac{1}{4}\left|\frac{\Delta}{U_0}\right| \to \frac{1}{2}, \text{ at } U_0 \to +\infty. \tag{34}$$

Thus, the conclusion we arrive at is that in the linear case under strong coupling conditions the population is approximately equally distributed (recall the normalization) between the two energy levels. This is a remarkable peculiarity of the Nikitin model since in the Landau–Zener case we have a "normal", intuitive limit $p_{LZ}(U_0^2 \to \infty) = 1$.

Consider now the nonlinear case. For the detuning modulation function (22) the limit equation (18) takes the form

$$\left[\Delta^2(1-e^{-at})^2 + 4U_0^2(1-3p)\right]p_t + \frac{U_0^2}{2}\frac{a}{(e^{at}-1)}(1-8p+12p^2) = 0. \tag{35}$$

The appropriate solution to this equation satisfying the initial condition $p(-\infty) = 0$ reads

$$p_0 = \frac{1}{6} + \frac{(1-e^{-at})}{18U_0^2/\Delta^2}\left((1-e^{-at}) + \sqrt{(1-e^{-at})^2 + \frac{6U_0^2}{\Delta^2}}\right). \tag{36}$$

This is a monotonically increasing function of time. The plots for $U_0^2/\Delta^2 = 3$ and 4 are shown in Fig. 4 together with the corresponding numerical solution of the initial equation (3). As is seen, the agreement, as a (uniformly valid) zero-order approximation, is good. Furthermore, note that the Rabi-type oscillations observed at $t > 0$ in the solution to the exact equation sharply vanish as the field intensity is increased so that they can be disregarded with high accuracy. Thus, under the strong-coupling conditions the derived solution well describes the process (being asymptotically exact).

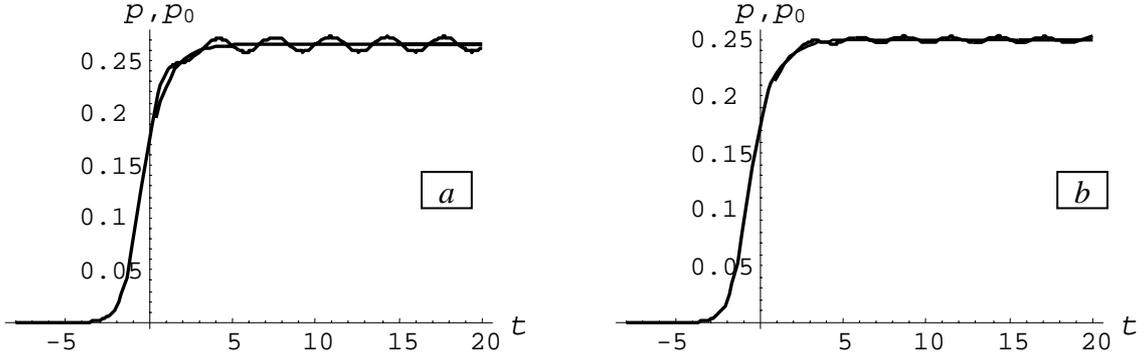

Fig. 4. Nikitin model: transition probability: (a) $U_0^2/\Delta^2 = 3$, (b) $U_0^2/\Delta^2 = 4$.
Monotonic curves correspond to solution (36).

Consider further the consequences following from this solution. The final transition probability at $t \to +\infty$ is given as

$$p_0(+\infty) = \frac{1}{6} + \frac{\Delta^2}{18 U_0^2}\left(1 + \sqrt{1 + \frac{6 U_0^2}{\Delta^2}}\right). \qquad (37)$$

Note now that, importantly, this formula does not involve the parameter $a$ that defines the resonance crossing rate in the Nikitin model (22). Hence, the final transition probability is not defined by the Landau–Zener parameter $\lambda_{LZ} = 2U_0^2/|a\Delta|$. Instead, the final conversion probability is a function of the Nikitin parameter $\lambda_N = U_0^2/\Delta^2$ which involves the final detuning $\Delta$. Hence, we see that the behavior of a system at term crossing is highly affected if a finite final detuning is involved.

Furthermore, in contrast to the Landau-Zener case, the probability (37) is a monotonically *decreasing* function of the field intensity $U_0^2$. Starting from $U_0^2/\Delta^2 = 4/3$, $p_0$ becomes less than $1/3$. As is seen, at $U_0^2 \to \infty$ the final probability tends, always being more than $1/6$, to the limit $1/6$ [compare with the linear result, Eq. (34)]:

$$p_0(+\infty)\big|_{U_0^2/\Delta^2 \to \infty} \sim \frac{1}{6} + \frac{1}{3\sqrt{6}}\left|\frac{\Delta}{U_0}\right| \to \frac{1}{6}. \qquad (38)$$

Thus, in the extreme limit of strong coupling only the one third ($= 2 \times 1/6$) of the atoms is converted into molecules. [Recall here that $1/6$ is one of the stationary solutions to Eq. (3)]. Obviously, this is a consequence of the *finite final detuning* $\Delta$ present in function (22) – the hallmark of the Nikitin model, that is to say the only essential difference from the Landau-Zener model. As we have seen above, this finite detuning suppresses the transition to the second state already in the linear case. However, it turns out that the interaction of the Rabi-type nearly periodic oscillations caused by this finite detuning with the nonlinear terms involved in Eq. (3) further suppresses the population of the molecular state as compared with the linear case. An immediate conclusion following from these observations is that the strong-coupling limit *is not optimal* for transition to the molecular state when dealing with models that involve, such as the above Nikitin model, a finite final detuning. Furthermore, since $p_0(+\infty)$ monotonically decreases as the field intensity is increased, for a given final detuning $\Delta$ an optimal field intensity $U_0^2 = U_0^2(\Delta)$ should exist for which the transition to the molecular state is maximal.

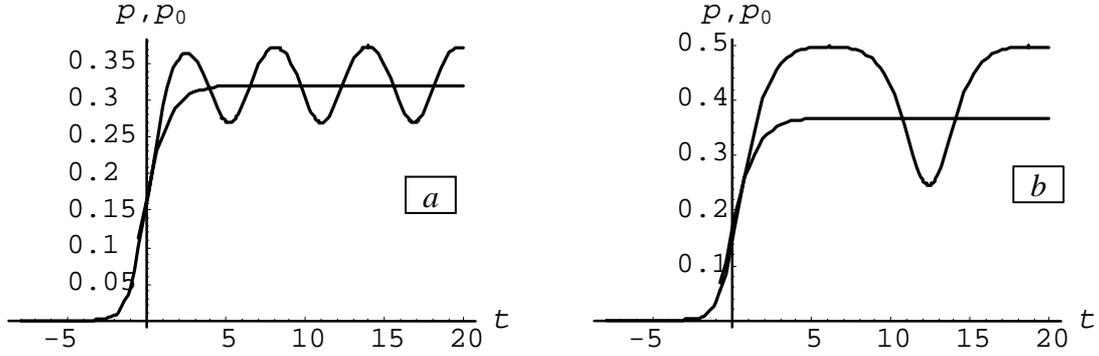

Fig. 5. Intermediate regime: (a) $U_0^2/\Delta^2 = 1.5$, (b) $U_0^2/\Delta^2 = 1$.
Nonoscillatory curves present limit solution (36).

This conclusion is confirmed by close examination of the intermediate regime of moderately strong couplings. The time evolution of the molecular state probability in this regime is shown in Fig. 5. As can be shown, in this case the first-order solution in the region $t \gg 1/a$ can be presented as a sum of the limit solution (36) and a slightly modified Rabi solution. Indeed, the linearization of Eq. (3) by means of the substitution $p = p_0 + u$, $p_0$ being the limit solution (36), leads to the following equation

$$u_{ttt} - \frac{a}{(e^{at}-1)} u_{tt} + \left[\Delta^2(1-e^{-at})^2 + 4U_0^2(1-3p_0)\right] u_t - 12U_0^2 p_{0t} u \qquad (39)$$
$$- 4U_0^2 \frac{a}{(e^{at}-1)}(1-3p_0)u + D(p_0) = 0,$$

with
$$D(p_0) = \left( p_{0ttt} - \frac{a}{(e^{at}-1)} p_{0tt} \right). \tag{40}$$

In the region $t \gg 1/a$, where $\delta_t \approx \Delta$ we have $p_0 \approx const$ and, as a result, Eq. (39) is reduced to the equation

$$u_{ttt} + \left[\Delta^2 + 4U_0^2(1-3p_0)\right] u_t = 0 \tag{41}$$

that has exactly the same form as the one describing the *linear* Rabi oscillations but with the parameter $U_0^2$ replaced by $U_0^2(1-3p_0)$. Note that at large $U_0^2/\Delta^2$ the latter, i.e., $U_0^2(1-3p_0)$, is approximately equal to $U_0^2/2$ (in the Landau-Zener case we have $-U_0^2/2$). The amplitude $A$ and the phase $\varphi_0$ of the oscillations,

$$u = C_0 + A\sin\left(\sqrt{\Delta^2 + 4U_0^2(1-3p_0)}\, t + \varphi_0\right), \tag{42}$$

should be defined from the solution of Eq. (39) for the region $t < 1/a$. This solution can be straightforwardly constructed asymptotically since the higher-order derivatives are small. The resultant solution (rather cumbersome) then shows that $C_0 = 0$ and that the amplitude of the Rabi oscillations decreases exponentially as $U_0^2/\Delta^2$ increases, becoming negligible, of the order of 1%, already at $U_0^2/\Delta^2 \approx 4$. At less field intensities, $U_0^2/\Delta^2 < 4$, as is seen from Fig. 5, we encounter strong Rabi oscillations, practically sinusoidal up to $U_0^2/\Delta^2 \approx 1.5$ and with well-pronounced nonlinear changes of the shape at $U_0^2/\Delta^2 \lesssim 1$. These observations allow us to state that the approximate optimum for the transition to the molecular state is achieved at $U_0^2/\Delta^2 \approx 4$ (for moderate $\Delta \sim 1$). If this optimum is chosen, the final transition probability is $\approx 1/4$, i.e., much more than $1/6$ given by the limit solution (36) at $U_0^2/\Delta^2 \to \infty$. Note that the limit $p_{+\infty} = 1/4$ corresponds to the nonreversible transformation to the molecular state of the half of the initial atomic population. The solution for this optimal regime, revealing just weakly pronounced Rabi-type oscillations, is the one shown in Fig. 4b.

## 5. Summary

In summary, we have presented an analysis of the strong nonlinearity regime for different term-crossing models for a nonlinear version of the two-state problem arising in photoassociation of an atomic Bose-Einstein condensate. The discussion is based on a third-order nonlinear differential equation for the molecular state probability derived from the initial set of coupled first-order equations for the probability amplitudes. We have presented a class of periodic level-crossing models permitting exact solution in terms of elliptic functions. The models reveal generic features such as, for instance, large-amplitude periodic oscillations that are common for the constant-detuning noncrossing Rabi problem.

Examining in general the single term-crossing case via variation of the optical field detuning, we have shown that in the limit of strong coupling, when the nonlinearity is most pronounced, the governing equations are effectively replaced by a first-order nonlinear ordinary differential equation. This limit equation has a rich structure and possesses several solutions. In general, the zero-order approximation for the time evolution of the transition probability in the strong interaction limit presents a function composed from different solutions to this equation. The limit solution allows linearizing the problem under consideration getting a linear third-order differential equation that well describes the behavior of the system everywhere.

Further, we have analyzed the first exponential term-crossing model by Nikitin comparing it with the Landau-Zener and Rabi models. We have shown that, because of a finite final detuning involved, in the limit of large laser field intensities the final transition probability for the Nikitin model tends to 1/6 while in the intermediate regime of moderate field amplitudes, where only slightly expressed, small-amplitude Rabi-type oscillations occur, the probability is about 1/4–1/3. Thus, the general conclusion is that the strict strong interaction limit, perhaps surprisingly, is not an optimal for molecule formation. We have found that the optimum for the transition to the molecular state is achieved at $U_0^2/\Delta^2 \approx 4$. If this optimum is chosen, the final transition probability is about $1/4$, i.e., almost the half of the population of the initial atomic condensate is converted into molecular state. This is the main physical result of the present paper.

Finally, we would like to mention that the presented procedure of description of the strong-interaction limit using the solution to the first-order limit equation as a zero-order approximation, since the general solution to the limit equation is known, can be a general approach for attacking analogous nonlinear curve crossing models such as the second Demkov-Kunike model and the double level-crossing model that are very helpful in numerous applications [27]. Also, we hope that similar developments will be possible in the case when other nonlinear processes, associated with, for instance, atom-atom, atom-molecule and molecule-molecule, interactions (see, e.g., [31]), are taken into consideration.

**Acknowledgments**

This work was supported by the International Science and Technology Center Grant no. A-1241 and by a Grant-in-Aid for Scientific Research on Specially Promoted Project no. 15002011 from the Ministry of Education, Culture, Sports, Science, and Technology of Japan.